\documentclass[preprint2]{aastex}

\begin{document}

\newcommand{\psra}{\mbox{PSR J0737$-$3039A}}
\newcommand{\psrb}{\mbox{PSR J0737$-$3039B}}

\title{The Mean Pulse Profile of \psra}
\author{R.\ N.\ Manchester\altaffilmark{1}, M.\ Kramer\altaffilmark{2}, 
A. Possenti\altaffilmark{3}, A. G. Lyne\altaffilmark{2},   \\
M.\ Burgay\altaffilmark{3}, I. H. Stairs\altaffilmark{4}, 
A.\ W.\ Hotan\altaffilmark{5}, M.\ A.\ McLaughlin\altaffilmark{2},   \\
D.\ R.\ Lorimer\altaffilmark{2}, G.\ B.\ Hobbs\altaffilmark{1}, 
J.\ M. Sarkissian\altaffilmark{1}, N.\ D'Amico\altaffilmark{6},  \\
F. Camilo\altaffilmark{7}, B. C. Joshi\altaffilmark{8} \& 
P. C. C. Freire\altaffilmark{9}}

\altaffiltext{1}{Australia Telescope National Facility, CSIRO,
P.O. Box 76, Epping NSW 1710, Australia}
\altaffiltext{2}{Jodrell Bank Observatory, University of Manchester, Macclesfield, 
Cheshire, SK11 9DL, UK}
\altaffiltext{3}{INAF - Osservatorio Astronomico di Cagliari, Loc. Poggio dei
Pini, Strada 54, 09012 Capoterra, Italy}
\altaffiltext{4}{Dept. of Physics and Astronomy, University of British
Columbia, 6224 Agricultural Road, Vancouver BC V6T 1Z1, Canada}
\altaffiltext{5}{Swinburne Centre for Astrophysics \& Supercomputing, Hawthorn, 
VIC 3122, Australia}
\altaffiltext{6}{Universit\`a degli Studi di
Cagliari, Dipartimento di Fisica, 09042 Monserrato, Italy}
\altaffiltext{7}{Columbia Astrophysics Laboratory, Columbia University, 
550 West 120th Street,  NY 10027}
\altaffiltext{8}{National Centre for Radio Astrophysics, P.O. Bag 3, Ganeshkhind, 
Pune 411007, India}
\altaffiltext{9}{NAIC, Arecibo Observatory, HC03 Box 53995, PR 00612}

\begin{abstract}
General relativity predicts that the spin axes of the pulsars in the
double-pulsar system (PSR J0737$-$3039A/B) will precess rapidly, in
general leading to a change in the observed pulse profiles. We have
observed this system over a one-year interval using the Parkes 64-m
radio telescope at three frequencies: 680, 1390 and 3030 MHz. These
data, combined with the short survey observation made two years
earlier, show no evidence for significant changes in the pulse profile
of \psra, the 22-ms pulsar. The limit on variations of the profile
10\% width is about $0\fdg5$ per year. These results imply an angle
$\delta$ between the pulsar spin axis and the orbit normal of
$\lesssim 60 \degr$, consistent with recent evolutionary studies of
the system. Although a wide range of system parameters remain
consistent with the data, the model proposed by \citet{jr04} can be
ruled out. A non-zero ellipticity for the radiation beam gives
slightly but not significantly improved fits to the data, so that a
circular beam describes the data equally well within the
uncertainties.
\end{abstract}

\keywords{pulsars: general --- pulsars: individual (J0737--3039A)}

\section{Introduction} \label{sec:intro}
The double-pulsar system PSR J0737$-$3039A/B \citep{bdp+03,lbk+04}
provides a wonderful laboratory for investigations of relativistic
gravity \citep{klb+04} as well as the physics of pulsar magnetospheres
\citep[e.g.,][]{mkl+04}. The system consists of a 22-ms pulsar, \psra,
in a 2.4-h binary orbit with \psrb, a younger pulsar with a spin
period of 2.7 s. The system is mildly eccentric ($e \sim 0.088$) and
is viewed nearly edge-on (orbit inclination $i \sim 88\degr$). With
mean orbital speeds $v \sim 0.001 c$, the system is highly
relativistic, allowing the detection of four ``post-Keplerian''
parameters in just six months of observation \citep{lbk+04}.  The
post-Keplerian parameters, together with the mass ratio, uniquely
measurable in this double-pulsar system, give accurate values for the
masses of the two stars as well as stringent tests of general
relativity \citep{klb+04}.

Within the framework of general relativity, the spin vectors of the
two pulsars are expected to exhibit geodetic precession about the
total angular momentum of the system \citep{dr74,bo75}. Since the
total angular momentum is dominated by the orbital motion, the pulsar
spin vectors effectively precess about the orbit normal. The
precession rate $\Omega$ depends on the Keplerian parameters and the
masses of both pulsars, with predicted precessional periods of $\sim
75$ yr and $\sim 71$ yr for \psra~and \psrb~ respectively
\citep{lbk+04}. These are about a factor of four shorter than the value
for PSR B1913+16, the Hulse-Taylor binary pulsar (Weisberg, Romani \&
Taylor 1989)\nocite{wrt89}. In general, precession will result in a
change of the width and shape of the observed profile as the angle of
the line of sight with respect to the pulsar spin axis ($\zeta$)
changes during the precessional period. Long-term evolution of the
mean pulse profile has been observed in PSR B1913+16
\citep{wrt89,kra98} and PSR B1534+12 \citep{sta04} and interpreted as
evidence for geodetic precession. This gives the three-dimensional
geometry (within a $180\degr$ ambiguity) of the system and, in the
case of PSR B1913+16, a prediction that the pulsar will no longer be
visible after the year 2025. \citet{wt02} have used the precessional
motion to produce a two-dimensional image of the part of the PSR
B1913+16 emission beam traversed so far, suggesting that the beam is
elongated in the latitudinal direction.

If the magnetic-pole model \citep{rc69a} is assumed, $\zeta$ and the
inclination angle of the magnetic axis with respect to the spin axis
($\alpha$) can be deduced from a fit of the model to the observed
variations of polarization position-angle, at least for wide profiles
\citep[e.g.,][]{lm88}. \citet{drb+04} used such an analysis to
conclude that \psra~is a nearly aligned system with $\alpha \sim
4\degr$, but the fit to the observed position angle variations is
poor. The observer angle $\zeta$ is unconstrained by this
fit. \citet{jr04} determined possible geometries of the
J0737$-$3039A/B system based on a model for the orbital modulation of
the \psrb~pulse intensity. Two solutions were obtained with
($\alpha,\delta$) of ($1\fdg6 \pm 1\fdg3, 167\degr \pm 10\degr$) and
($14\degr \pm 2\degr, 90\degr \pm 10\degr$) respectively. The first of
these is consistent with the \citet{drb+04} solution. Both solutions
predict a rapid evolution of the observed profile width of \psra, with
expected changes of $\sim 42\degr$ and $\sim 96\degr$ per year for the
two models. Here we report on observations of the mean pulse profile
of \psra~over a three-year interval, 2001 August to 2004 August.

\section{Observations and Analysis} \label{sec:obs}
We have used the Parkes 64-m radio telescope to observe the PSR
J0737$-$3039A/B system since 2003 May 1 (MJD 52670), shortly after its
confirmation, to 2004 August 8 (MJD 53225). Observations were made in
three frequency bands centered at 680, 1390 and 3030 MHz, with the
680-MHz (50cm) and 3030-MHz (10cm) observations commencing in 2003,
December. Before 2003 October, the 1390 MHz observations were made
using the center beam of the multibeam receiver, which has an
equivalent system noise of approximately 29 Jy. In addition, we have
the original survey pointing, a 4-min observation made at 1390 MHz on
2001 August 22 \citep{bdp+03}. After 2003 October, the H-OH receiver
with a system noise of about 42 Jy was used. Both the multibeam and
H-OH systems were dual polarization and used a filterbank system
consisting of $2 \times 512 \times 0.5$ MHz channels. After detection,
the two polarizations for each channel were summed, sampled at either
80 $\mu$s or 125 $\mu$s intervals and recorded to tape for subsequent
processing. The other two bands were observed using the
dual-polarization coaxial 10cm/50cm receiver, with system noises of
$\sim 48$ Jy and $\sim 64$ Jy at 10cm and 50cm respectively. A
filterbank system with $2 \times 256 \times 0.25$ MHz channels was
used with the 50cm receiver and at 10cm the filterbank system had $2
\times 192 \times 3$ MHz channels.

\begin{figure}[ht]
\includegraphics[scale=0.35,angle=270]{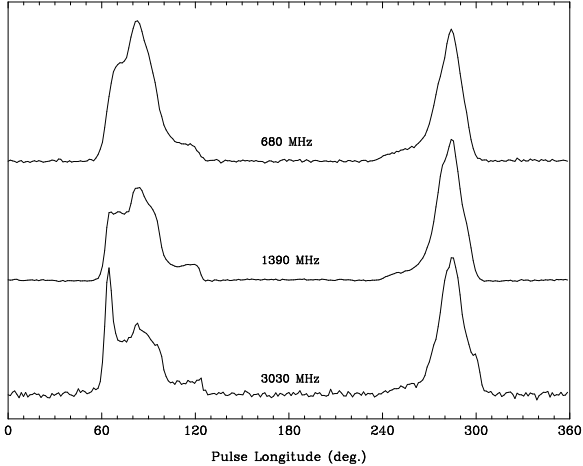}
\caption{Mean pulse profiles for \psra~at three frequencies, where
  $360\degr$ of longitude corresponds to the pulse period. The 680-MHz
  and 3030-MHz profiles were aligned to give maximum cross-correlation
  with the 1390-MHz profile. }\label{fg:3frq}
\end{figure}

Over the 15-month interval, we obtained 59 observations at 1390 MHz,
17 observations at 680 MHz and 27 at 3030 MHz, all with durations
ranging from 10 min to 5 hr. Data from each observation
were folded at the apparent topocentric period of \psra~to form mean
pulse profiles with 256 phase bins, which were then summed in
frequency and time to form a single profile using the PSRCHIVE data
analysis system (Hotan, van Straten \& Manchester
2004)\nocite{hvm04}. All observations for each frequency band were
then summed to form grand-average profiles containing 116, 27 and 34
hrs of data, respectively. These profiles are shown in
Figure~\ref{fg:3frq}.

\begin{figure}[ht]
\includegraphics[scale=0.45]{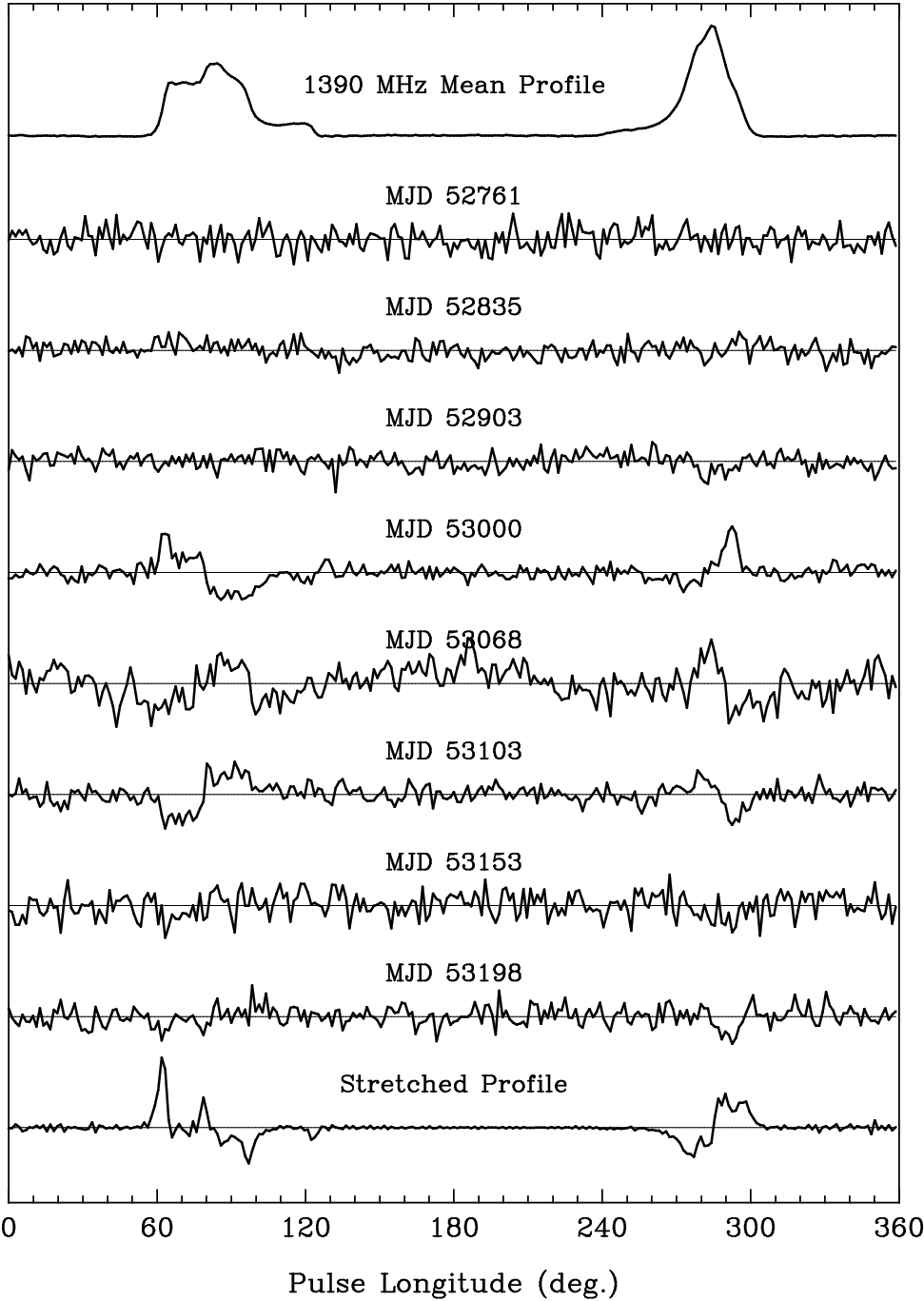}
\caption{Difference profiles for \psra~at 1390~MHz relative to the
grand-average pulse profile shown at the top of the figure for eight
epochs, together with the difference profile corresponding to the
grand-average profile stretched by 0.4\%. The vertical scale of the
difference profiles is a factor of ten larger than that of the
grand-average profile.}\label{fg:diff}
\end{figure}

The 1390-MHz observations over the 15 months from 2003 May were
divided into eight chronologically-ordered groups and average profiles
were formed for each of these eight groups. Integration times ranged
between 5 hrs (MJD 53068) and 20 hrs (MJD 53000). Examination of these
average profiles showed little or no evidence for evolution of the
mean pulse profile over the 15-month interval. The eight profiles were
each aligned with the grand-average profile using a cross-correlation
analysis and baselines subtracted, taking the longitude range
$310\degr$ to $50\degr$ (Figure~\ref{fg:3frq}) as the baseline
region. Each profile was scaled to have the same area as the
grand-average profile, which was then subtracted from it to form the
difference profiles shown in Figure~\ref{fg:diff}. Also shown at the
bottom of the figure is the difference profile corresponding to a
stretched version of the grand-average profile, demonstrating the
signature of a change in profile width. The adopted stretch factor
(1.004) corresponds to an increase of about $1\degr$ in profile width,
where the width is defined to be the separation of the steep outer
edges of the profile (at roughly $60\degr$~and $300\degr$~in
Figures~\ref{fg:3frq} and \ref{fg:diff}) at 10\% of the amplitude of
the second (stronger) component.

Significant differences are seen, especially between MJDs of 53000 and
53103. In general, these differences do not have the signature of an
increase or decrease in profile width, but rather appear to result
from changes in the relative amplitude of the various pulse
components, especially at the leading and trailing components of the
profile. Polarization observations \citep[e.g.,][]{drb+04} show that
these components have high linear polarization. It is likely that the
observed changes are an instrumental effect resulting from differences
in receiver gain between the two polarization channels combined with
parallactic angle variations. 

These difference profiles confirm the lack of significant secular
profile evolution. There is no evidence for a systematic change in the
overall profile width to a level of much less than $1\degr$ and there
is no evidence for any intrinsic changes in the width or shape of the
leading and trailing pulse components considered separately. We have
also reprocessed the original 4-min survey pointing to give a profile
with signal-to-noise ratio of about 25. Again, this shows no evidence
for any significant change in profile shape or width.

\begin{figure}[ht]
\includegraphics[scale=0.3,angle=270]{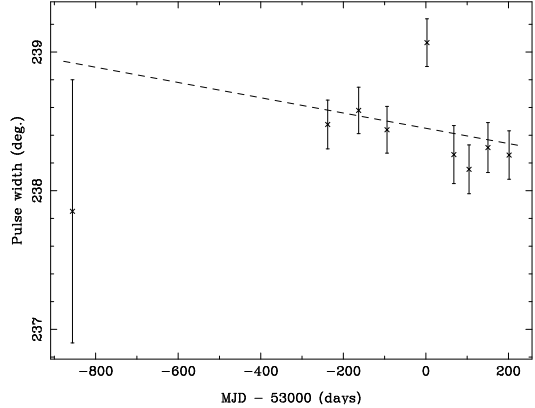}
\caption{Profile widths at 10\% of the peak amplitude for \psra~at
1390 MHz as a function of MJD relative to 53000. Error bars are $\pm 1
\sigma$. The dashed line is a formal weighted least-squares fit of a linear
trend to these data.}\label{fg:width}
\end{figure}

In order to quantify the profile stability, widths at 10\% of the
pulse peak were measured for each of the eight 1390-MHz profiles and
the 2001 observation using a cross-correlation technique.  The leading
($50\degr$ -- $130\degr$ in Figures~\ref{fg:3frq} and \ref{fg:diff})
and trailing ($240\degr$ -- $310\degr$) pulse components were
separately cross-correlated with a standard profile derived from the
grand-average profile containing just the relevant component,
determining the phase of each component relative to a reference
phase. The difference of these phases was then added to the 10\% width
of the grand-average profile to give the 10\% width at each
epoch. Errors from the cross-correlation analysis were increased in
quadrature by 10 $\mu$s to allow for systematic effects. The derived
10\% widths are plotted in Figure~\ref{fg:width} along with the fitted
trend line. Clearly, there is no significant trend, with the fitted
slope being $-0\fdg2 \pm 0\fdg3$ per year. The 10\% profile width at
MJD 53000 (2003 December 27) from the fit is $238\fdg45 \pm
0\fdg12$. Quoted errors are $\pm 2\sigma$ for both the width and
slope.

\section{Discussion and Conclusions} \label{sec:discn}
Figure~\ref{fg:3frq} confirms the overall bilateral symmetry of the
\psra~pulse profile, consistent with its origin from magnetic field
lines associated with a single magnetic pole
\citep{bdp+03,drb+04}. There are significant differences in the
spectral indices of the various pulse components. The most striking of
these is the relatively flat spectrum of the component at the extreme
leading edge of the profile which is relatively much stronger at 3030
MHz. The 3030-MHz profile also shows a matching flat-spectrum
component at the extreme trailing edge of the profile, reinforcing the
idea that both components of the observed profile originate
on field lines associated with a single magnetic pole. We assume this
in the following discussion.

The main result of this paper is the extraordinary stability of the
\psra~pulse profile over more than three years, a significant fraction
of the expected geodetic precessional period of 75 years. Provided the
angle between the spin axis of the pulsar and the orbital angular
momentum ($\delta$) is non-zero, we would expect to see a variation in
profile width as a function of time. In particular, the system
geometries proposed by \citet{jr04} predict changes of many tens of
degrees per year in the observed profile width. Clearly, these are not
observed, ruling out this model in its present form.

One possible explanation for the observed lack of variation in
profile width is that the spin axis of \psra~is aligned with the
orbital angular momentum, i.e., $\delta \sim 0$. Since we view the
orbit nearly edge-on, even for an orthogonal magnetic axis ($\alpha
\sim 90\degr$), this would require an effective beam radius $\rho >
90\degr$, which seems unlikely. Furthermore, given that \psrb~probably
suffered a significant natal kick, an aligned rotation axis for \psra~
is possible but unlikely (Willems, Kalogera \& Henninger
2004)\nocite{wkh04b}.

\begin{figure}[ht]
\includegraphics[scale=0.35,angle=270]{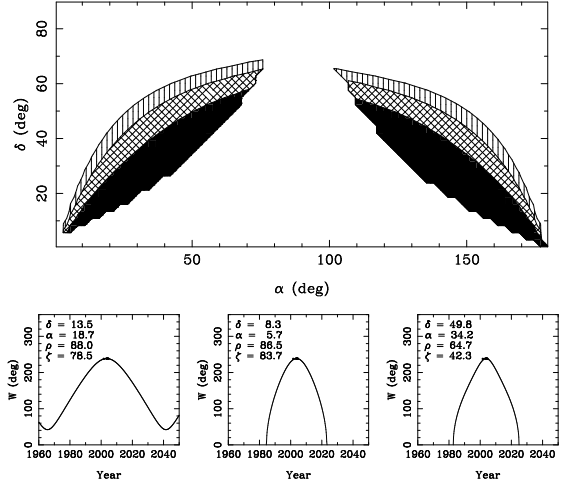}
\caption{Results of fits to the observed profile widths of a model for
  geodetic precession of the pulsar spin axis, inclined by an angle
  $\delta$ to the orbit normal, with a circular beam of radius $\rho$
  inclined at an angle $\alpha$ to the pulsar spin axis. The upper
  part shows contours of $\chi^2$ at 1-$\sigma$ intervals, relative to
  the minimum value, with the darker regions having lower
  $\chi^2$. The lower cutoff to the allowed regions represents the
  locus of points where the beam radius $\rho = 90\degr$. These
  contours are mirrored for $90\degr < \delta < 180\degr$ with
  ($\alpha$, $\delta$) corresponding to ($180\degr - \alpha$, $180\degr
  - \delta$). In the lower part of the figure, the variation of
  profile width is shown as a function of time for three cases: (a) a
  point near the global minimum for $\chi^2$, (b) a small $\alpha$,
  small $\delta$ case, and (c) a solution with intermediate $\alpha$
  and $\delta$. The angle $\zeta$ is the minimum value of the
  inclination of the line of sight to the pulsar spin axis which
  occurs at precessional phase zero (or $180\degr$) when the pulsar
  spin axis is in the plane defined by the orbit normal and the line
  of sight.}\label{fg:prec}
\end{figure}

We have investigated the limits which can be placed on the geometry of
the system by modeling the observed profile width as a function of
time \citep[cf.][]{kra98}. An inclination angle $i = 88\degr$
is assumed\footnote{Since only $\sin i$ is determined at present,
equivalent solutions are possible with $i=92\degr$ and $\alpha
\rightarrow 180\degr-\alpha$, etc.}.  For a circular emission beam,
four parameters are sufficient to model the expected profile changes:
$\alpha$, $\delta$, $\rho$ and an epoch $T_0$ describing the
precession phase. We fitted a circular beam model with $\rho<90\degr$
to the data and derive a $\chi^2$-sphere plot for $\alpha$ and
$\delta$ as shown in Figure~\ref{fg:prec}. This figure shows that
solutions with misalignment angle $\delta \lesssim 60\degr$ are
preferred, consistent with the conclusions of \citet{wkh04b} based on
the observed velocity of the system.

The formal best-fit solution (reduced $\chi^2 = 2.88$) is located at
$\alpha\sim 19\degr$ and $\delta \sim 14\degr$ (and at a corresponding
mirror-solution of $\alpha\sim 161\degr$ and $\delta \sim
166\degr$). The predicted time variation for the observed profile width
for this configuration is shown in Figure~\ref{fg:prec}(a); pulses
would be detected over the whole precessional period with an
approximately sinusodial variation in the profile width. However, the
$\chi^2$-sphere is rather flat, and a statistically satisfactory
solution is possible over a wide range of
angles. Figure~\ref{fg:prec}(b) shows a solution with small $\alpha$
and $\delta$ which is consistent with the preferred solution of
\citet{drb+04} based on the observed position angle variations. Both
of these solutions have a beam radius of very close to $90\degr$
corresponding to a fan beam. In case (b) however, the pulse is visible
for a limited period only, roughly 1984 to 2023. Figure~\ref{fg:prec}(c)
shows a solution with intermediate $\alpha$ and $\delta$ for which the
beam radius is smaller, about $65\degr$. In all solutions, the
observed pulse is emitted from field lines associated with a single
magnetic pole as indicated by the observed profile symmetry.

Unless $\delta\sim0$, the essentially constant profile width observed
so far implies that the system is at or close to precessional phase
zero (or $180\degr$). At these times is ${\rm d}W/{\rm d}t \sim
0$.\footnote{There is a special case with $\cos\rho\sin\delta\cos\Phi
= \cos\alpha$, where $\Phi$ is the precessional phase, which results in
${\rm d}W/{\rm d}t = 0$, but this requires an improbable fine-tuning
of the system parameters.}  This is possible -- a similar situation
evidently applies for PSR B1913+16 \citep{kra98} -- but statistically
unlikely. A possible way of relaxing this constraint is to allow a
non-circular beam.  \citet{big90b} and \citet{ks98} have shown that,
under certain assumptions, the polar emission beam is compressed in
the latitudinal direction for $\alpha>0\degr$. On the other hand,
polarization observations of young and millisecond pulsars
\citep[e.g.,][]{nv83,mh04} suggest beams effectively elongated in the
latitude direction. First, we took the model of \citet{ks98} in which
the beam compression is a function of $\alpha$ only and recomputed the
$\chi^2$-sphere. The results were only marginally different to those
shown in Figure~\ref{fg:prec}. We then took the beam ellipticity as a
free parameter, first doing a grid search in ellipticity at each point
on the ($\alpha$,$\delta$) plane and then searching for a global
$\chi^2$ minimum in ($\rho_m$, $T_0$, $\epsilon$) space, where
$\rho_m$ is the beam major axis, $T_0$ is the time of precessional
phase zero and $\epsilon$ is the beam ellipticity.  Latitudinally
compressed beams with axial ratio $\rho_{\rm lat}/\rho_{\rm long}
\lesssim 0.5$ are restricted to a fairly small range of $\alpha$ and
$\delta$ around $45\degr$. On the other hand, fan-like beams with
large axial ratio allow a much larger range of $\alpha$ and $\delta$
with values $\lesssim 30\degr$ preferred. However, in none of these
cases was the $\chi^2$ value significantly better than those found in
the circular-beam case.

In conclusion, we find that the pulse-width variations (or lack of
them) observed so far allow a wide range of configurations for the PSR
J0737$-$3039A/B system. Models with non-circular beams give somewhat
better fits to the data but are statistically indistinguishable from
fits with a circular beam. Misalignment angles $\delta \lesssim
60\degr$ are generally preferred, consistent with the conclusions of
\citet{wkh04b}, and the configurations discussed by \citet{jr04}
can be ruled out. Solutions in which the pulse never disappears from
view are possible, as are solutions where it does, in the most extreme
cases about 15 years from now. Clearly, a longer time baseline, even
one more year, will help to constrain the models, as will a more
realistic interpretation of the observed position-angle variations.

\acknowledgments

The Parkes radio telescope is part of the Australia Telescope which is
funded by the Commonwealth Government for operation as a National
Facility managed by CSIRO. IHS holds an NSERC UFA and is supported by
a Discovery Grant. DRL is a University Research Fellow funded by the
Royal Society.  FC acknowledges support from NSF grant
AST-02-05853. NDA, AP and MB received support from the Italian
Ministry of University and Research (MIUR) under the national program
{\it Cofin 2003}.

%\bibliographystyle{apj} 
%\bibliography{journals,modrefs,psrrefs,crossrefs}

\end{document}